\documentstyle[12pt]{article}

\def\be{\begin{equation}\begin{array}{c}}
\def\ee{\end{array}\end{equation}}
\def\lae{{\lambda\over\eta}}
\begin{document}
\begin{titlepage}

\begin{flushright}
ITEP-TH-4/97
\end{flushright}
\vskip 1in
\begin{center}
{\LARGE Symmetries of BFKL Equation}\\
\bigskip\bigskip
{\Large A.Mikhailov}\\
\bigskip
Institute of Theoretical and Experimental Physics\\
117259, Bol. Cheremushkinskaya, 25, Moscow, Russia\\
and\\
Department of Physics, Princeton University\\
Princeton, NJ 08544, USA
\end{center}
\vskip 1in
\begin{abstract}
We discuss the algebraic structure of the spin chains
related to high energy scattering in QCD. We study
the sl(2) Yangian symmetry and possible generalizations
to nonzero spin and anisotropy parameter.
\end{abstract}
\end{titlepage}

\section{Introduction.}

Since seventies, much evidence was obtained that the high
energy scattering of hadrons can be described by exactly
solvable model. It describes interaction
of Reggeons -- the collective coordinates, which appear
in QCD in the Regge limit. The corresponding Hamiltonian
was discovered in the
paper \cite{KLF-76}.
The authors studied the high energy scattering of two vector
particles in the theory with
spontaneously broken gauge symmetry.

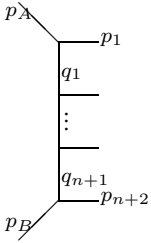
\begin{picture}(100,100)
\put(5,5){\line(1,1){15}}
\put(0,10){$\scriptstyle p_B$}
\put(5,95){\line(1,-1){15}}
\put(0,90){$\scriptstyle p_A$}
\put(20,20){\line(0,1){60}}
\put(20,20){\line(1,0){15}}\put(36,20){$\scriptstyle p_{n+2}$}
\put(21,27){$\scriptstyle q_{n+1}$}
\put(20,40){\line(1,0){15}}
\multiput(23,47)(0,3){3}{\circle*{1}}
\put(20,60){\line(1,0){15}}
\put(21,67){$\scriptstyle q_1$}
\put(20,80){\line(1,0){15}}\put(36,80){$\scriptstyle p_1$}
\end{picture}

\vspace{-100pt}

\hangindent=100pt  \hangafter=-8    By
means of unitarity, the scattering amplitude  can be related to the
integral  of the  product  of   two  $2\rightarrow  2+n$   amplitudes
over  the   momenta of intermediate $2+n$ states,  and sum over $n$.
The regge limit  is when   $s=(p_A+p_B)^2\gg   m^2$   and
$-(p_A-p_B)^2\sim   m^2$.   In this limit the main contribution from
the integral over the intermediate states comes from the so called
multiregge region, where

\be
1\gg -(p_B,q_1)/s\gg
-(p_B,q_2)/s\gg\ldots\gg -(p_B,q_{n+1})/s \sim m^2/s\;,     \\ m^2/s\sim
(p_A,q_1)/s\ll\ldots\ll (p_A,q_n)/s\ll (p_A,q_{n+1})/s\ll 1
\ee

One of the key observations was the specific form
of the $2\rightarrow 2+n$ amplitude in this limit:

\begin{equation}\label{multiregge}
\begin{array}{c}
A_{2\rightarrow 2+n}=
s\frac{(s_1/m^2)^{\alpha(q_1^2)}}{(q_1^2-m^2)}
\cdots \frac{(s_{n+1}/m^2)^{\alpha(q_{n+1}^2)}}
{(q_{n+1}^2-m^2)}\\
\Gamma_{AD_0}^{i_1} \gamma_{i_1i_2}^{D_1}\cdots
\gamma_{i_ni_{n+1}}^{D_n} \Gamma_{BD_{n+1}}^{i_{n+1}}
\end{array}
\end{equation}

where
\be
\alpha(k)=-{g^2\over (2\pi)^3}\int d^2q
\frac%
{|k|^2+m^2}
{(|q-k|^2+m^2)(|q|^2+m^2)}=-{g^2\over 8\pi^2}
\log\left({|k|^2\over m^2}\right)+o(m^2)
\ee

is related to the Regge trajectory $j=1+\alpha(t)$ and

\begin{equation}
\gamma_{ij}=m\delta_{ij},\;\;
\gamma_{ij}^D(q_1,q_2)=ig\epsilon_{ijD} {\cal P}_{\mu}(q_1,q_2) e_D^{\mu}
\end{equation}

are the effective vertices \cite{KLF-76}
for emission of scalar and vector particles.
If the particle $D$ is on shell,
{\em i.e.} $(q_1-q_2)^2=m^2$, the effective vertex equals:

\begin{equation}
\begin{array}{rl}
{\cal P}^{\mu}(q_1,q_2)=-(q_1+q_2)^{\mu}_{\perp}&
-\alpha_1 p_A^{\mu}\left(1-2{|q_{1\perp}|^2+m^2\over
|q_{1\perp}-q_{2\perp}|^2+m^2}\right)\\
 &
 +\beta_2 p_B^{\mu}\left(1-2{|q_{2\perp}|^2+m^2\over
|q_{1\perp}-q_{2\perp}|^2+m^2}\right)
\end{array}
\end{equation}

The important property of the ${\cal P}_{\mu}$ is that if
the emitted particle is on shell, the scalar product of the
vertices:

\begin{equation}
\sum_D \gamma_{ij}^D(k_1,p_1)\gamma_{i'j'}^D(k_2,p_2)=
2g^2 (K^{(0)} c_{ij}c_{i'j'} +
      K^{(1)} c_{ij}^k c_{i'j'}^k +
      K^{(2)} c_{ij}^{kl} c_{i'j'}^{kl})
\end{equation}

depends only on the components of momenta,
orthogonal to the plane $(p_A,p_B)$. Here the coefficients $c$ are
the projectors on the space with definite isospin. In particular,
the component of isospin $0$ is:

\begin{equation}
\begin{array}{c}
K^{(0)}=-2|k_1+k_2|^2-3m^2
+2\frac{(|k_1|^2+m^2)(|p_2|^2+m^2)+(|p_1|^2+m^2)(|k_2|^2+m^2)}
{|k_1-p_1|^2+m^2}
\end{array}
\end{equation}

This expression in the limit $m\to 0$ goes to:

\be
K^{(0)}\stackrel{m\to0}{=}{2\over |q|^2}(k_1k_2^*p_1^*p_2+\mbox{c.c.})
\ee

Thus, the sum of the ladder diagrams \footnote{In the LLA, the main contribution to the integral
over the intermediate states comes from the multiregge region.}:

\begin{center}
\begin{picture}(250,50)
\put(10,10){\line(1,0){40}}\put(10,13){$\scriptstyle q_0$}
\put(10,40){\line(1,0){40}}\put(10,33){$\scriptstyle p_0$}
\put(60,20){$+$}
\put(80,10){\line(1,0){40}}
\put(80,13){$\scriptstyle q_0$}\put(115,13){$\scriptstyle q_1$}
\put(80,40){\line(1,0){40}}
\put(80,33){$\scriptstyle p_0$}\put(115,33){$\scriptstyle p_1$}
\put(100,10){\line(0,1){30}}\put(101,26){$\scriptstyle k_1$}
\put(130,20){$+$}
\put(150,10){\line(1,0){40}}
\put(150,40){\line(1,0){40}}
\put(165,10){\line(0,1){30}}
\put(175,10){\line(0,1){30}}
\put(195,20){$+$}
\put(210,20){$\cdots$}
\put(235,20){$=$}
\end{picture}
\end{center}

\be\label{BS}
=\phi_s(q_0,p_0)=\sum\limits_{n=0}^{\infty}
\left({N_c g^2\over (2\pi)^3}\right)^{n}
{(s_0/m^2)^{\alpha(q_0)+\alpha(p_0)}\over (|q_0|^2+m^2)(|p_0|^2+m^2)}
\times\\\times
\prod_{j=1}^n\left[\int d^2k_{j\perp}\int {d\alpha_j\over\alpha_j}
\frac%
{2[q_jp_j^*q_{j-1}^*p_{j-1}+\mbox{h.c.}]}
{|k_j|^2+m^2}
\frac%
{(s_{j}/m^2)^{\alpha(q_j)+\alpha(p_j)}}
{(|q_j|^2+m^2)(|p_j|^2+m^2)}
\right]
\ee

is described by the BFKL integral
equation.
Here $q_j=q_{j-1}-k_j$, $p_j=p_{j-1}+k_j$.
The mass $m$ plays the role of regulator. $N_c$ is the number of colors.
We denoted $q_j$, $p_j$ the $\perp$-components
of the momenta $q^{\mu}_j$ and $p^{\mu}_j$, and the
$\|$-components of $q^{\mu}_j$ are expressed through the
Sudakov's variables:
$$
q_{j\|}=\alpha_j p_A - \beta_j p_B
$$

$|q|^2$ means $-q_{\perp}^2$. The integration is over the momenta
of the intermediate particles, which are on shell.
Notice that (in the Regge limit):
\be
s_j=(q^{\mu}_{j-1}-q^{\mu}_{j+1})^2=
-(\alpha_{j-1}-\alpha_{j+1})(\beta_{j-1}-\beta_{j+1})s-\\
-|q_{j-1}-q_{j+1}|^2
=
\alpha_{j-1}\beta_{j+1}s-|q_{j-1}-q_{j+1}|^2
\ee

On the other hand, from the mass-shell condition for the
emitted particle
%
follows $\beta_js={1\over \alpha_{j-1}}(m^2+|q_{j-1}-q_j|^2)$.
Thus
$
s_j={\alpha_{j-1}\over\alpha_j}(m^2+|q_j-q_{j+1}|^2)
$

In particular,
\be\label{product}
\prod\limits_{j=0}^n {s_j\over m^2}={1\over \alpha_n}
\prod\limits_{j=0}^n \left(1+{|q_j-q_{j+1}|^2\over m^2}\right)
=
{s\over m^2}
\left(1+{|k_1|^2\over m^2}\right)
\left(1+{|k_2|^2\over m^2}\right)
\cdots
\left(1+{|k_n|^2\over m^2}\right)
\ee

In the LLA, the product $(1+|k_n|^2/m^2)$ may be omitted.
The formula ${s\over m^2}=\prod {s_i\over m^2}$ enables
to interpret $\log{s\over m^2}$ as time in the corresponding integrable
system: $\tau_j=\log(s_j/m^2)$. We can also write the integration
measure $\prod_j{d\alpha_j\over\alpha_j}$ as $\prod_j d\tau_j$.
Thus, the equation (\ref{BS}) can be written as:
\be
\phi_{\tau}(q,p)=\sum\limits_{n=0}^{\infty}
\left[{N_c g^2\over (2\pi)^3}\right]^{n} 2^n
{\int\limits_{\tau>\tau_n>\ldots>\tau_1>0} d\tau_n\cdots d\tau_1}
e^{[\alpha(q)+\alpha(p)](\tau-\tau_n)}\times\\\times
\prod\limits_{j=1}^n\left\{ \left[
{1\over q^*p}\int{d^2k_j\over |k_j|^2+m^2}
e^{k_j(\partial_p-\partial_q)+k_j^*(\partial^*_p-\partial^*_q)}
q^*p+\mbox{h.c.}\right]e^{[\alpha(q)+\alpha(p)](\tau_j-\tau_{j-1})}
\right\}{1\over |q|^2|p|^2}
\ee

If we denote $ix=\partial_p$, $iy=\partial_q$, and take the integrals
over $dk_j$, we get:
\be
\phi_{\tau}(q,p)=\exp\left(-\tau {N_c g^2\over 8\pi^2}\hat{H}\right)
\frac%
{N_c g^2/(2\pi)^3}
{|q|^2|p|^2}
\ee

where\footnote{Here we followed the paper
\cite{KLF-76}. See \cite{KLS} for another derivation, using
the lagrangian language.}
\be\label{HolAntihol}
\hat{H}=\log(|q|^2)+\log(|p|^2)+
{1\over qp^*}\log(|x-y|^2) qp^* +
{1\over q^*p}\log(|x-y|^2) q^*p - 4\psi(1)
\ee

This hamiltonian is the sum of two expressions, acting on
holomorphic and antiholomorphic functions.

The part of the scattering amplitude, corresponding to the
exchange of pomeron, will contain the factor
$\exp\left[-h {N_c g^2\over 8\pi^2}\log s\right]$ where $h$ is the
corresponding eigenvalue of $\hat{H}$.







In this paper we discuss the algebraic properties of the
Hamiltonian (\ref{HolAntihol}).
We review the derivation of this Hamiltonian from the transfer-matrix.
We will show, how it commutes with the $sl(2)$ Yangian, and
discuss its various forms.
In the last section we will consider the integrable perturbation
of this Hamiltonian, which on the language of spin chains
corresponds to turning on small anisotropy.

\section{Spin chains.}
We want to construct an integrable spin chain such that the
spin-$j$ representation of $sl(2)$ is associated with each site.
This system will have an infinite number of commuting
hamiltonians. In the framework of Quantum Inverse Scattering Method,
these hamiltonians are included into the noncommutative algebra.
This algebra is generated by the entries of the monodromy matrix
$T(\lambda)$, which acts in the auxiliary vector space.
The commutation relations are encoded in the $RTT$ equations:
\be
R(\lambda-\mu)T(\lambda)T(\mu)=T(\mu)T(\lambda)R(\lambda-\mu)
\ee

Where $R$ is the (numerical) $R$-matrix. An important property
of such algebras is the existence of "comultiplication". Namely,
if we have two matrices $T_1(\lambda)$, $T_2(\lambda)$, satisfying
the $RTT$ relation, then their product
$T_1(\lambda)\stackrel{.}{\otimes}T_2(\lambda)$ also satisfies
these relations. Schematically, if

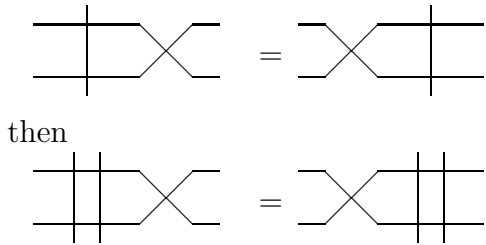
\begin{picture}(200,40)
\put(10,10){\line(1,0){40}}
\put(10,30){\line(1,0){40}}
\put(50,10){\line(1,1){20}}
\put(50,30){\line(1,-1){20}}
\put(70,10){\line(1,0){10}}
\put(70,30){\line(1,0){10}}
\put(30,3){\line(0,1){34}}
\put(95,15){$=$}
\put(110,10){\line(1,0){10}}
\put(110,30){\line(1,0){10}}
\put(120,10){\line(1,1){20}}
\put(120,30){\line(1,-1){20}}
\put(140,10){\line(1,0){40}}
\put(140,30){\line(1,0){40}}
\put(160,3){\line(0,1){34}}
\end{picture}

then

\begin{picture}(200,40)
\put(10,10){\line(1,0){40}}
\put(10,30){\line(1,0){40}}
\put(50,10){\line(1,1){20}}
\put(50,30){\line(1,-1){20}}
\put(70,10){\line(1,0){10}}
\put(70,30){\line(1,0){10}}
\put(25,3){\line(0,1){34}}
\put(35,3){\line(0,1){34}}
\put(95,15){$=$}
\put(110,10){\line(1,0){10}}
\put(110,30){\line(1,0){10}}
\put(120,10){\line(1,1){20}}
\put(120,30){\line(1,-1){20}}
\put(140,10){\line(1,0){40}}
\put(140,30){\line(1,0){40}}
\put(155,3){\line(0,1){34}}
\put(165,3){\line(0,1){34}}
\end{picture}

This suggests the way to construct the monodromy matrix and the hamiltonians:
one finds some simplest representation of $RTT$ -algebra (which corresponds
to one site of the chain), and then constructs the $T$-operator for the
whole chain using the comultiplication.

For example, the $RTT$ relations are satisfied by the entries of
the $2\times 2$ matrix
\be
L_{elem.}(\lambda)=1+\sum\limits_{a=1}^3 {1\over\lambda}\sigma_a S^a
\ee

where $S^a$ are the generators of $sl(2)$ in the spin-$j$
representation. Taking the product of these "elementary"
$L$-operators over all the sites, we get some $2\times 2$
matrix, whose entries are the operators acting in the
Hilbert space of the system. The traces of this matrix will
commute for different values of the spectral parameter.
Unfortunately, the trace of $L$, considered as the power
series in the spectral parameter, will have nonlocal operators as
coefficients, thus we cannot consider them as hamiltonians.

The way out is to consider more complicated object, "the
fundamental $L$-operator", associated with each site.
The fundamental $L$-operator is the matrix in $2j+1$-dimensional
"auxiliary" space, whose entries are the operators acting in
the local space on the site. We need something like
$$
L=\lambda+\sum\limits_{a=-j}^j S^a_{aux}S^a_q \eqno (wrong)
$$

but such a simple $L$ would not satisfy any $RTT$ relation,
and more care is needed to construct them correctly.
The fundamental $L$-operators formally coincide with the
fundamental $R$-matrix intertwining them.

The $RTT$ relations ensure the integrability of the system
\cite{BIKS,FK,K,WJ}.

In the next section we review the construction of the
fundamental $R$-matrix and derive the hamiltonian of the
solvable spin chain. Actually we will consider the case of XXZ
model, where the symmetry is not $sl(2)$, but $U_q(sl(2))$,
the quantum algebra. The hamiltonian depends on
$q$ -- the deformation parameter.

\section{Fusion procedure. The fundamental R-matrix.}

The idea \cite{TTF,KRS,JLMP}
is to consider first the $R$-matrix for the spin-$1/2$ chain,
study the symmetry properties of this $R$-matrix and then to construct
the spin-$j$ $R$-matrix from it. Remember that the spin-1/2 matrix
is related to the 6-vertex model statistical weights,
\be
\left[
\begin{array}{cccc}
\sin(\lambda+\eta)&0&0&0\\
0&\sin(\lambda)&\sin(\eta)&0\\
0&\sin(\eta)&\sin(\lambda)&0\\
0&0&0&\sin(\lambda+\eta)
\end{array}
\right]
\ee

where $\eta$ is anisotropy parameter. Introduce $z=e^{i\lambda}$,
$q=e^{i\eta}$, and consider the modified $2\times 2$ R-matrix,
which differs from the original one by conjugation:

\be
R'_{4\times 4}(z)=z^{\frac{1}{2}\sigma^3\otimes 1}
R(z)z^{-\frac{1}{2}\sigma^3\otimes 1}=\\
=\left[
\begin{array}{cccc}
zq-{1\over zq}& 0 &0 &0\\
0 & z-{1\over z} & z(q-{1\over q}) &0\\
0&{1\over z}(q-{1\over q}) & z-{1\over z} &0\\
0&0&0&zq-{1\over zq}
\end{array}
\right]
\ee

The  purpose of this conjugation is to ensure that the
$4\times 4$ $R$-matrix satisfies the following symmetry properties:

\begin{equation}\label{Inter1}
\begin{array}{c}
(q^{\sigma^3\over 2}\otimes \sigma^{\pm}+\sigma^{\pm}\otimes
q^{-{\sigma^3\over 2}})R'(z)=\\= R'(z)(q^{-{\sigma^3\over 2}}
\otimes\sigma^{\pm}+\sigma^{\pm}\otimes q^{\sigma^3\over 2})
\end{array}
\end{equation}

and

\begin{equation}\label{Inter2}
\begin{array}{c}
\left( q^{-{\sigma^3\over 2}}\otimes {\sigma^+\over z}+
{\sigma^+\over w}\otimes q^{\sigma^3\over 2}\right)
R'\left(\sqrt{z\over w}\right)=\\=
R'\left(\sqrt{z\over w}\right)
\left({\sigma^+\over w}\otimes
q^{-{\sigma^3\over 2}}+q^{\sigma^3\over 2}\otimes
{\sigma^+\over z}\right)
\end{array}
\end{equation}

The first of these properties means, that the
$R$-matrix is invariant under the quantum group ${\cal U}_q(sl(2))$.
The second extends this symmetry to
${\cal U}_q(\widehat{sl}(2))$,
-- the quantum Kac-Moody algebra. (See Appendix.)

Let us show that there are $R$-matrices acting in the tensor
product of spin-$j$ ($j>1/2$) representations, which satisfy the
similar symmetry properties. Indeed, we may construst the $R$-matrix
for, say, spin -1, by means of the following "fusion procedure":

\begin{picture}(220,100)
\put(5,50){$R'\left({z\over w}\right)=$}
\put(105,0){\line(0,1){100}}\put(93,0){\small -2}
\put(180,0){\line(0,1){100}}\put(168,0){\small -1}
\put(65,20){\line(1,0){140}}\put(200,10){\small 1}
\put(65,80){\line(1,0){140}}\put(200,70){\small 2}
\put(140,25){$\scriptstyle R'_{4\times 4}\left({qz\over w}\right)$}
\put(140,85){$\scriptstyle R'_{4\times 4}\left({z\over w}\right)$}
\put(60,25){$\scriptstyle R'_{4\times 4}\left({z\over w}\right)$}
\put(60,85){$\scriptstyle R'_{4\times 4}\left({z\over qw}\right)$}
\end{picture}

(The matrices are multiplied from the bottom right to the top
left.) Notice the special relations between the arguments: they differ
by multiplication on the power of $q$. The reason is that we want the
tensor products of $R$'s to act correctly on the symmetric part
of the tensor product of two-dimensional spaces. For example,
consider the product
$R'\left({z\over w}\right)\otimes
 R'\left({qz\over w}\right)$
 (the two rightmost R's on the figure). They act on the tensor
 product $1\otimes 2$. Notice that the "q-symmetric" tensors\footnote{
 q-symmetric are the tensors which become symmetric after action of
 $q^{\frac{1}{2}\sigma^3\otimes 1}$}
  remain q-symmetric.
 Indeed, the antisymmetrization $A_{12}$ of $1\otimes 2$ can be expressed
through the $R$-matrix:
$A^{(q)}_{12}=q^{-\frac{1}{2}\sigma^3\otimes 1}A_{12}
q^{\frac{1}{2}\sigma^3\otimes 1}={1\over q^{-1}-q}R'(q^{-1})$
Thus, using the Yang-Baxter relations
$R_{12}R_{13}R_{23}=R_{23}R_{13}R_{12}$, we have:
\be
A^{(q)}_{12} R'_{2,-1}\left({z\over w}\right)
             R'_{1,-1}\left(q{z\over w}\right)
           =\\=
             R'_{1,-1}\left(q{z\over w}\right)
             R'_{2,-1}\left({z\over w}\right) A^{(q)}_{12}
\ee

so the space of q-symmetric tensors is preserved.
The matrix $R'\left({z\over w}\right)$ acts in the tensor product
$$
\mbox{Symm}_q ({\bf C}^2)^{\otimes n}\otimes
\mbox{Symm}_q ({\bf C}^2)^{\otimes n}
$$

The space $\mbox{Symm}_q ({\bf C}^2)^{\otimes n}$ is an irreducible
representation of quantum $sl(2)$, and it also can be considered as
the irreducible part of
$$
\rho^{1/2}_{z^2}\otimes\rho^{1/2}_{q^2 z^2}\otimes\cdots
\otimes\rho^{1/2}_{q^{2(n-1)}z^2}
$$

as described in Appendix.
This yields the desired generalization of the symmetry properties of
fundamental R-matrix:

\begin{equation}\label{ordinarySL2}
\begin{array}{c}
\left[PR'\left(\sqrt{z\over w}\right),
\left(q^{-S_3}\otimes E+E\otimes q^{S_3}\right)
\right]=0
\end{array}
\end{equation}

and
\begin{equation}\label{extrasymmetry}
\begin{array}{c}
PR'\left(\sqrt{z\over w}\right)
\left(q^{S_3}\otimes {E\over z}+{E\over w}\otimes q^{-S_3}
\right)=\\=
\left(q^{S_3}\otimes {E\over w}+{E\over z}\otimes q^{-S_3}
\right)
PR'\left(\sqrt{z\over w}\right)
\end{array}
\end{equation}

Notice that the Yang-Baxter equations for $R'$ can be derived from
the explicit construction from the Yang-Baxter equations for
$4\times 4$ R-matrix, as well as from the consistency of these
symmetry properties.

From (\ref{ordinarySL2}) we infer that:
\begin{equation}
PR'\left(\sqrt{z\over w}\right) =
\sum\limits_{p=0}^{n}
F_p(z/w)P_{p}
\end{equation}

where $P_p$ is the projector on the subspace in the tensor product,
which has spin $p$.

 The coefficients $F_p$ may be found from
(\ref{extrasymmetry}). 

For example, one may consider the action
of (\ref{extrasymmetry}) on the highest weight vector:
\be
\left|\left({n\over 2},{n\over 2}\right)_p,p\right\rangle=
{1\over\sqrt{(n-p)_q(p+1)_q}}\left|{n\over 2},{n\over 2}\right\rangle
\otimes\left|{n\over 2},p-{n\over 2}\right\rangle
-\\-
{q^{-p-1}\over\sqrt{(n)_q}}\left|{n\over 2},{n\over 2}-1\right\rangle
\otimes \left|{n\over 2},p-{n\over 2}+1\right\rangle+\ldots
\ee

The key point is that the operator
$\left(q^{S_3}\otimes {E\over z}+
{E\over w}\otimes q^{-S_3}\right)$ is the tensor one -- it follows
from the Serre relations, see Appendix. Thus, when we act by this
operator on
$\left|\left({n\over 2},{n\over 2}\right)_p,p\right\rangle$,
we get spins not higher than $p+1$. The coefficient of
$\left|\left({n\over 2},{n\over 2}\right)_{p+1},p+1\right\rangle$
can be found from

\begin{equation}\label{expression}
\begin{array}{c}
\left(q^{S_3}\otimes {E\over z}+ {E\over w}\otimes q^{-S_3}
\right)\left|\left({n\over 2},{n\over 2}\right)_p,p\right\rangle
=\\=
\left({q^{n/2}\over z}-
{q^{-2p-2+{n\over 2}}\over w}\right)
\left|{n\over 2},{n\over 2}\right\rangle\otimes
\left|{n\over 2},p-{n\over 2}+1\right\rangle+\ldots
\end{array}
\end{equation}

We get:
\be
F_{p+1}(z/w)\left({q^{n/2}\over z}-
{q^{-2p-2+{n\over 2}}\over w}\right)=
F_p(z/w)\left({q^{n/2}\over w}-
{q^{-2p-2+{n\over 2}}\over z}\right)
\ee

-- the relation between $F_p$ and $F_{p+1}$. Finally, we get the
expression for $R'$:

\begin{equation}
\begin{array}{c}
PR'(z)=P_0-P_1q^{-2}
\frac{1-z^2q^2}{1-z^2q^{-2}}+
P_2q^{-6}\frac{1-z^2q^2}{1-z^2q^{-2}}
\frac{1-z^2q^4}{1-z^2q^{-4}}-\ldots
\end{array}
\end{equation}

This expression can be rewritten in terms of Gamma-functions,
which enables to generalize it for general (non-integer) $j$
\footnote{We consider the product of two spin-$j$ representations.
The capital $J$ denotes the operator $\sum pF_p$.}:
\be
PR'(e^{i\lambda})=(-1)^J q^{-J(J+1)}
\frac{\Gamma_{q^2}(\lae+J+1)\Gamma_{q^2}(\lae-J)}
{\Gamma_{q^2}(\lae+1)\Gamma_{q^2}(\lae)}
\ee

$PR'(1)=1$.
The rational limit corresponds to taking $\lambda$ and $\eta$ very
small and keeping their ratio finite.

The hamiltonian of the integrable model is related to the R-matrix:

\be\label{HviaR}
R(z=1+\lambda)=P+\lambda PH+o(\lambda)
\ee

where the symmetric $R$-matrix is related to the $R$-matrix which
we constructed in fusion procedure by conjugation:
\be
R(z)=g R'(z) g^{-1},\\
g=\prod\limits_{k<0}z^{-\sigma_k^3\over 2}
  \prod\limits_{k<0}q^{-(n+k+1){\sigma_k^3\over 2}}
  \prod\limits_{k>0}q^{-k{\sigma_k^3\over 2}}
\ee

-- the numeration of the vector spaces as on the figure.
Notice that the last two products in $g$,
$\prod\limits_{k<0}q^{-(n+k+1){\sigma_k^3\over 2}}
 \prod\limits_{k>0}q^{-k{\sigma_k^3\over 2}}$
are symmetric (commute with $P$). They ensure that $R$ acts on
symmetric tensors (not q-symmetric). The first product,
$g_1=\prod\limits_{k<0}z^{-\sigma_k^3\over 2}$, ensures that
$R(z)$ is symmetric. One can see it from the fusion procedure, or
from the symmetry properties, eqs. (\ref{ordinarySL2}),
(\ref{extrasymmetry}). Indeed, conjugating these equations by
$g_1$, we get:
\be
PR\left(\sqrt{z\over w}\right)
\left(q^{-S_3}\otimes\sqrt{z}E +\sqrt{w}E\otimes q^{S_3}\right)
=\\=
\left(q^{-S_3}\otimes\sqrt{w}E +\sqrt{z}E\otimes q^{S_3}\right)
PR\left(\sqrt{z\over w}\right)
\ee

and
\be
PR\left(\sqrt{z\over w}\right)
\left(q^{S_3}\otimes\sqrt{w}E +\sqrt{z}E\otimes q^{-S_3}\right)
=\\=
\left(q^{S_3}\otimes\sqrt{z}E +\sqrt{w}E\otimes q^{-S_3}\right)
PR\left(\sqrt{z\over w}\right)
\ee

-- these equations are related to each other by permutation of
tensor multipliers. If $R$ were not symmetric, we could
construct the symmetric solution as $R+PRP$. But these equations
determine $R$ uniquely. Thus, $R$ is symmetric.

From (\ref{HviaR}) we get the hamiltonian:
\be\label{NSH}
P+\lambda PH+o(\lambda)= P+
\lambda\left(\sum\limits_{k>0}{\sigma_k^3\over 2}-
             \sum\limits_{k<0}{\sigma_k^3\over 2}\right)P
+\\+
{\lambda\over\eta}P(\psi_{q^2}(J+1)+\psi_{q^2}(-J))+o(\lambda)
\ee

We know from construction, that this expression is symmetric.
The hamiltonian is:
\be\label{HviaPsi}
H=\eta\left(\sum\limits_{k<0}{\sigma_k^3\over 2}-
        \sum\limits_{k>0}{\sigma_k^3\over 2}\right)+
        (\psi_{q^2}(J+1)+\psi_{q^2}(-J))
 =\\=
{1\over 2}\left(\psi_{q^2}(J+1)+\psi_{q^2}(-J)+
               P\psi_{q^2}(J+1)P+P\psi_{q^2}(-J)P\right)
\ee

And following the tradition in quantum group theory,
we denote

\be
\hbar=i \eta
\ee

\section{The rational case: Yangian symmetry.}

Here we consider separately the case $\hbar=0$. In this case instead
of the quantum Kac-Moody symmetry,
we have the Yangian symmetry \cite{Bernard}.
It can be derived from the basic property of the
fundamental $R$-matrix: it intertwines $L$-operators.
Namely, the entries of the $2\times 2$ matrix
\be
L(\lambda)=1+{1\over \lambda}\vec{\sigma}\cdot\vec{S}
\ee

satisfy the commutation relations of $RTT$ algebra:
\be
L_1(\lambda)L_2(\lambda+\mu)R_{12}(\mu)=
R_{12}(\mu)L_2(\lambda+\mu)L_1(\lambda)
\ee

Consider this equation as a series in $\lambda^{-1}$.
The zeroth order term is trivial, the first order term expresses
the fact that $R$ is $sl(2)$-invariant. And the coefficient
of $1\over \lambda^2$ is:
\be
(\vec{\sigma}\cdot\vec{S}_1)(\vec{\sigma}\cdot\vec{S}_2)
R_{12}(\mu)-\mu(\vec{\sigma}\cdot\vec{S}_2)R_{12}(\mu)
=\\=
R_{12}(\mu)
(\vec{\sigma}\cdot\vec{S}_2)(\vec{\sigma}\cdot\vec{S}_1)-
\mu R_{12}(\mu)(\vec{\sigma}\cdot\vec{S}_2)
\ee

Consider this equation up to first order in $\mu$,
$ R_{12}(\mu)=P_{12}+\mu P_{12}H_{12}+\ldots $:
\be
i[[\vec{S_1}\times\vec{S_2}],H_{12}]=\vec{S}_2-\vec{S}_1
\ee

Here $H_{12}$ is the two-site hamiltonian. If we had a chain with
$n$ sites, with the hamiltonian
\be
\sum\limits_{j=1}^n H_{j,j+1}
\ee

then we would get:
\be
[\sum\limits_{i<j} \vec{S}_i\times\vec{S}_j,
 \sum\limits_{j=1}^n H_{j,j+1}]= \vec{S}_n-\vec{S}_1
\ee

The operators
$\vec{\cal T}=\sum\limits_{i<j} \vec{S}_i\times\vec{S}_j$
commute with the hamiltonian modulo the boundary
terms\footnote{One cannot define $\vec{\cal T}$ for the periodic
chain, since the {\em linear} order is required, not just a
cyclic one.}. Together with the generators of $sl(2)$ they
generate the algebra called Yangian.

We want to construct an integrable spin chain with arbitrary
spin in the sites. We will use the functional representation of
$sl(2)$:

\be
S^+=z^2\partial_z-2jz,\\
S^3=z\partial_z-j,\\
S^-=-\partial_z
\ee

In this representation:
\be
{\cal T}^-=\sum\limits_{i<j}[(z_i-z_j)\partial_i\partial_j+
                            j(\partial_i-\partial_j)]
\ee

and the other ${\cal T}$'s may be obtained from ${\cal T}^-$ by
commutation with $sl(2)$.

Let us construct the Hamiltonian which respects these symmetries.
It is useful to go to the Fourier transform
\be
f(z)=\int e^{ikz}f(k)dk
\ee

and write an ansatz for the hamiltonian as:
\be
H.\phi(p_1,p_2)=\int dk_1 dk_2 D(p_1,p_2;k_1,k_2) \phi(k_1,k_2)
\ee

Let us require that the commutator with ${\cal T}_-$ is a local
operator. Then in the bulk of the integration region:
\be\label{YMS}
\left[
k_1k_2\left({\partial\over\partial k_1}-{\partial\over\partial k_2}
\right)+
p_1p_2\left({\partial\over\partial p_1}-{\partial\over\partial p_2}
\right)+\right.\\+\left.
(j+1)(p_1-p_2)-j(k_1-k_2)
\right] D(p_1,p_2;k_1,k_2)=0
\ee

%
%

The general solution of this equation is
\be
D(p_1,p_2;k_1,k_2)=\frac{(k_1k_2)^j}{(p_1p_2)^{j+1}}
D_0\left({k_2p_1\over k_1p_2};k_1+k_2,p_1+p_2\right)
\ee

Now we use the $sl(2)$-symmetry. From the commutation with
$s^-$ we infer, that $D_0$ should contain
$\delta(p_1+p_2-k_1-k_2)$. From the commutation with $s^3$
we know how $D_0$ scales with momenta, and find the dependence
on $k_1+k_2$:
$$
D_0=(k_1+k_2)\delta(p_1+p_2-k_1-k_2)
\tilde{D}_0\left({k_2p_1\over k_1p_2}\right)
$$

To find $\tilde{D}_0$, we have to consider the commutation
with $s^+$. We have:
\be
H\phi(p_1,p_2)=(p_1+p_2)\int dk
\frac{(p_1-k)^j(p_2+k)^j}{p_1^{j+1}p_2^{j+1}}
\tilde{D}_0\left({1+k/p_2\over 1-k/p_1}\right)
\phi(p_1-k,p_2+k)
=\\=
\int dk {1\over k}(1-k/p_1)^{j+1/2}(1+k/p_2)^{j+1/2}
F\left({1+k/p_2\over 1-k/p_1}\right)\phi(p_1-k,p_2+k)
\ee

where we have put $F(z)=(z^{1/2}+z^{-1/2})\tilde{D}_0$.
Acting on this integral by
\be
s^+=p_1{\partial^2\over\partial p_1^2}+
    p_2{\partial^2\over\partial p_2^2}+
    2(j+1)\left({\partial\over\partial p_1}+
                {\partial\over\partial p_2}\right)
\ee

and integrating by parts in $\int dk$, we get an equation for
$f$:
\be
\left[p_1{\partial^2\over\partial p_1^2}+
      p_2{\partial^2\over\partial p_2^2}-
      {1\over k}
      \left(p_1{\partial\over\partial p_1}+
            p_2{\partial\over\partial p_2}+1\right)
      \left(p_1{\partial\over\partial p_1}-
            p_2{\partial\over\partial p_2}\right)
+\right.\\ \left.+
      2(j+1)\left({\partial\over\partial p_1}+
                  {\partial\over\partial p_2}\right)\right]
(1-k/p_1)^{j+1/2}(1+k/p_2)^{j+1/2}
F\left({1+k/p_2\over 1-k/p_1}\right)=0
\ee

or, introducing $x=1-k/p_1$ and $y=1+k/p_2$:
\be
\left[-x(1-x)^2{\partial^2\over\partial x^2}
      +y(1-y)^2{\partial^2\over\partial y^2}+\right.\\ \left.
      +2j\left((1-x)^2{\partial\over\partial x}-
               (1-y)^2{\partial\over\partial y}\right)\right]
(xy)^{j+1/2}F(x/y)=0
\ee

or
\be
\partial_x\partial_y [(xy)^{j+1/2}F(x/y)]=0
\ee

which means that either \[(xy)^{j+1/2}F(x/y)=x^{2j+1}\] or
\[(xy)^{j+1/2}F(x/y)=y^{2j+1}\]

Since the hamiltonian should be symmetric, we have to add
these solutions:
\be
H.\phi(p_1,p_2)=\int {dk\over k}\left[
\left(1-{k\over p_1}\right)^{1+2j}+
\left(1+{k\over p_2}\right)^{1+2j} \right]
\phi(p_1-k,p_2+k)
\ee

Introducing $ix_{12}=\partial/\partial p_1-\partial/\partial p_2$, we
get:
\be
H=\int {dk\over k}\left[ {1\over p_1^{1+2j}}e^{-ikx_{12}}p_1^{1+2j}+
                   {1\over p_2^{1+2j}}e^{-ikx_{12}}p_2^{1+2j}\right]
\ee

The integral over $k$ is divergent and the cutoff is
needed. From the $s^3$ invariance we see, that the
cutoff should be of the form $\epsilon p_1$ or
$\epsilon p_2$. The corresponding hamiltonian will be
\be
H=\log(p_1p_2)+{1\over p_1^{1+2j}}\log(x_1-x_2)p_1^{1+2j}+
               {1\over p_2^{1+2j}}\log(x_1-x_2)p_2^{1+2j}
\ee

Indeed,
\be
\left[
z_1^2{\partial\over\partial z_1}+z_2^2{\partial\over\partial z_2}
-2j(z_1+z_2), H\right]=0
\ee

and
\be
\left[(z_1-z_2)\partial_1\partial_2-j(\partial_2-\partial_1),H
\right]=\partial_2-\partial_1
\ee

Notice that naively from the equation (\ref{YMS}) it follows
that the Hamiltonian exactly commutes with ${\cal T}^-$.
The origin of the boundary terms in commutation relation with
Yangian generator ${\cal T}^-$ is an infrared divergency in the Fourier
integral. That's why the RHS has such a simple form, local in the
momentum space. It turns out that this divergency does not
affect commutation with $s^+$ \cite{BarePomeron}.

Let us illustrate how to derive this hamiltonian (in a slightly
different form) directly from (\ref{HviaPsi}). Let us first
restrict ourselves with the case $j=0$.
From the representation
for $\psi$-function:
\be
\psi(z)=-\sum_{s=0}^{\infty} {1\over s+z}+\mbox{const}
\ee

we get:
\be
H=\sum\limits_{l=0}^{\infty}{2l+1\over l(l+1)+
                             (z_1-z_2)^2\partial_1\partial_2}
=\\=
\sum\limits_{l=0}^{\infty}{2l+1\over l(l+1)+
     x_1^2\left({\partial^2\over\partial x_0^2}
               -{\partial^2\over\partial x_1^2}\right)}
\ee

where $x_0=z_1+z_2$ and $x_1=z_1-z_2$. This expression is equal to
\be
H(\vec{n}\cdot\vec{n}')=4\pi \sum\limits_{l=0}^{\infty}
x_1\cdot\frac{\sum\limits_{m=-l}^lY^*_{lm}(\vec{n}')Y_{lm}(\vec{n})}
{{l(l+1)\over x_1^2}-{\partial^2\over\partial x_1^2}-
{2\over x_1^{\phantom 1}}
{\partial\over\partial x_1^{\phantom 1}}+
{\partial^2\over\partial x_0^2}}\cdot{1\over {\displaystyle x_1^3}}
\ee

when $\vec{n}=\vec{n}'$. The last expression is the Green
function for the Laplace operator in 4 dimensions,
\be
rG(t,t';r,r';\vec{n},\vec{n}')(r')^{-3}
=\\=
{r\over (r')^3}
\frac{\delta(|\vec{r}-\vec{r}'|-t+t')+
      \delta(|\vec{r}-\vec{r}'|+t-t')}{2|\vec{r}-\vec{r}'|}
\ee

The expression ${1\over |x-y|}$ , properly regularized, is the
kernel of the operator
$$
-2\log\partial-2\log z
$$

This gives us the second form of the hamiltonian:
\be
H=2\log (z_1-z_2)+
(z_1-z_2)\log(\partial_1\partial_2) (z_1-z_2)^{-1}
\ee

From the explicit expression for $J(J+1)$:
\be
J(J+1)=-(z_1-z_2)^2\partial_1\partial_2
-2j(z_1-z_2)(\partial_1-\partial_2)+2j(2j+1)
\ee

we see, that $J(J+1)$ for $j\neq 0$ is conjugate to $J(J+1)$
for $j=0$,
and consequently the hamiltonians for different spins are
conjugate:
\be
H_J=(z_1-z_2)^{2j} H_0 (z_1-z_2)^{-2j}
\ee

Finally, let us write down two forms of hamiltonian for  general
$j$:
\be\label{TwoForms}
H_j=\log(\partial_1\partial_2)+
{1\over \partial_1^{1+2j}}\log(z_1-z_2)\partial_1^{1+2j}+
{1\over \partial_2^{1+2j}}\log(z_1-z_2)\partial_2^{1+2j}
=\\=
2\log(z_1-z_2)+(z_1-z_2)^{1+2j}\log(\partial_1\partial_2)
               (z_1-z_2)^{-1-2j}
\ee

The Hamiltonian for general $J$ may correspond to taking into account
some logarithmic corrections. The similar (but essencially
different) Hamiltonian was considered in \cite{Kirschner}.

\section{Trigonometric case for small $\hbar$.}
In this section we will calculate the hamiltonian
in the anisotropic ($q\neq 1$) case to first nontrivial order
in $\hbar$ -- the deformation parameter.
We will restrict ourselves with spin zero.
Remember that the Hamiltonian is

\be
{1\over 2}\left(\psi_{q^2}(-J)+\psi_{q^2}(J+1)+(\hbar\to -\hbar)
\right)
\ee

The operator $J$ is the quantum Casimir operator in the product of
two representations. These two representations are spin-0
representations of quantum group in the space of functions,
whose arguments we will denote $z$ and $w$. The ${\cal U}_q(sl(2))$
operators act as follows:
\be
t=q^{-2d}\\
e={1\over x}{q^d-q^{-d}\over q-q^{-1}}\\
f=x {q^{-d}-q^d\over q-q^{-1}}
\ee

(we denoted $d=x\partial_x$, $x=$ $z$ or $w$.) The Casimir operator
for the quantum group:
\be
q^{2J_q+1}+q^{-(2J_q+1)}={1\over 2}\left[(q-q^{-1})^2(ef+fe)+
(q+q^{-1})(t+t^{-1})\right]
\ee

is an element of the center of ${\cal U}_q(sl(2))$. In the product of
two representations, it acts as:
\be\label{QuantumCasimir}
q^{2J_q+1}+q^{-(2J_q+1)}=\\=q+q^{-1}+
\left[\left({qw\over z}+{z\over qw}\right)-
(q+q^{-1})\right](t_z-1)(t_w^{-1}-1)
\ee

or, to the fourth order in $\hbar$:
\be
2+\hbar^2(1+4J(J+1))+\hbar^3[-4(d_z-d_w)J(J+1)-8{\cal T}^3]
+\\+
\hbar^4\left[{1\over 12}
-J(J+1)[4d_zd_w-{8\over 3}(d_z^2+d_w^2)-2]-
8{\cal T}^3(d_z-d_w)\right]+\ldots
\ee

where we used the Yangian generator
${\cal T}^3={1\over 2}(z^2-w^2)\partial_z\partial_w$.

The operator (\ref{QuantumCasimir})
has the same spectrum as $q^{2J+1}+q^{-2J-1}$, where $J$
is the classial momentum.
It is natural to conjecture that these operators are conjugate.
Indeed, we found to the order $\hbar^4$:
\be\label{conjugate}
q^{2J_q+1}+q^{-2J_q-1}=U (q^{2J+1}+q^{-2J-1}) U^{-1}
\ee

where
\be
U=\exp\left[{\hbar\over 2}(z^2-w^2)\partial_z\partial_w-
            {\hbar^2\over 12}(z^2-w^2)\partial_z\partial_w
            (d_z-d_w)+o(\hbar^2)\right]
\ee

Now
\be
H={1\over 2}[U(\psi_{q^2}(-J)+\psi_{q^2}(J+1))U^{-1}+
(\hbar\to -\hbar)]
\ee

Notice that the coefficient of $\hbar$ in
$\psi_{q^2}(-J_q)+\psi_{q^2}(J_q+1)$
equals to the commutator
\be
\left[{1\over 2}(z^2-w^2)\partial_z\partial_w,H_0\right]
=\\=
[{\cal T}^3,H_0]=
\left(w{\partial\over\partial w}-z{\partial\over\partial z}\right)
\ee

in agreement with eq. (\ref{HviaPsi}). After adding the permuted
(or $\hbar\to -\hbar$) expression, this term cancels.

Let us find the coefficient of $\hbar^2$ in $H$.
Because of the property (\ref{conjugate}),
it remains to calculate the expression
$\psi_{q^2}(-J)+\psi_{q^2}(J+1)$
where $J$ is the classical $J$, to the order $\hbar^2$.

Let us collect here a few properties of the quantum $\psi$-function.
Consider the function:
\be
f_q(z)=\sum\limits_{n=1}^{\infty}{z^n\over 1-q^n}=
       \sum\limits_{m=0}^{\infty}{zq^m\over 1-zq^m}
\ee

Write
\be
{1\over 1-q^n}=-{1\over \log q}
\left.{\partial\over\partial\epsilon}\right|_{\epsilon=0}
\frac{(q^{\epsilon};q)_n}{(q;q)_n}
\ee

where  $(a;q)_n=(1-a)(1-qa)\cdots (1-q^{n-1}a)$, and use the
"quantum binomial formula"
\be
\sum\limits_{n=0}^{\infty}{(a;q)_n\over (q;q)_n} z^n=
{(az;q)_{\infty}\over (z;q)_{\infty}}
\ee

and the definition of the $\Gamma$-function:
\be
\Gamma_q(x)={(a;q)_{\infty}\over (q^x;q)_{\infty}}
            (1-q)^{1-x}
\ee

We have :
\be
f_q(z)={\log(1-q)\over\log q}+{1\over \log q}\psi_q(x)
\ee

Consider the function (\cite{GasperRahman}, Ex. 1.21):
\be
{d\over dx}\psi_q(x)=(\log q)^2
\sum\limits_{n=0}^{\infty}\frac{q^{n+x}}{(1-q^{n+x})^2}=\\=
\sum_{n=0}^{\infty}\frac{\hbar^2}{2\cosh((n+x)\hbar)-2}
\ee

Take the derivative of this function with respect to $\hbar$:
\be
{\partial\over\partial\hbar}
\sum_{n=0}^{\infty}\frac{\hbar^2}{2\cosh((n+x)\hbar)-2}=\\=
\sum\limits_{n=0}^{\infty}\hbar
\left((n+x){\partial\over\partial (n+x)}+2\right)
{1\over 2\cosh\hbar(n+x)-2}
=\\=
\int_{\hbar x}^{\infty}dz (z\partial_z+2) {1\over 2\cosh z-2}+
{\hbar\over 2}\left(x{d\over dx}+2\right) {1\over 2\cosh \hbar x-2}
\ee

where we have used the formula
\be
\hbar \sum\limits_{n=0}^{\infty} f(x+n\hbar)=\int_x^{\infty} f(z)dz
+{\hbar\over 2} f(x)+ o(\hbar)
\ee

Taking the integral, we have:
\be
\psi_{q^2}(-J)+\psi_{q^2}(J+1)=\psi(-J)+\psi(J+1)+
{\hbar^2\over 3}J(J+1)
\ee

and
\be
\psi_{q^2}(-J_q)+\psi_{q^2}(J_q+1)=H_0+\\+
{\hbar\over 2}[(z^2-w^2)\partial_z\partial_w,H_0]+\\+
{\hbar^2\over 8}[(z^2-w^2)\partial_z\partial_w,
              [(z^2-w^2)\partial_z\partial_w,H_0]]-\\-
{\hbar^2\over 12}[(z^2-w^2)\partial_z\partial_w (d_z-d_w), H_0]-\\-
{\hbar^2\over 3}(z-w)^2\partial_z\partial_w
\ee

where we denoted by $H_0$ the XXX hamiltonian (\ref{TwoForms}).

\be
{[} (z-w)\partial_z\partial_w,H_0 {]} =\partial_w-\partial_z,\\
{[} (z^2-w^2)\partial_z\partial_w,H_0 {]} =2(d_w-d_z)
\ee

\underline{The terms of the second order in $\hbar$:}

We get the term
$$
{\hbar^2\over 2}(z^2+w^2)\partial_z\partial_w
$$
from the double commutator.

We also get the term ${\hbar^2\over 6}(d_z-d_w)^2$.

The most complicated expression arises from

\be
(z^2-w^2)\partial_z\partial_w [(d_z-d_w), H_0]
\ee

We have:
\be
H_0=\log(\partial_z\partial_w)+
{1\over \partial_z}\log(z-w)\partial_z+
{1\over \partial_w}\log(z-w)\partial_w
\ee

and
\be
(z^2-w^2)\partial_z\partial_w[(d_z-d_w),H_0]=\\=
(z^2-w^2)\partial_z\partial_w\left[
{1\over \partial_z}{z+w\over z-w}\partial_z+
{1\over \partial_w}{z+w\over z-w}\partial_w\right]
=\\=
(z^2-w^2)\left[
\partial_w{z+w\over z-w}\partial_z+
\partial_z{z+w\over z-w}\partial_w\right]
\ee

So we have:
\be
H=H_0+\hbar^2\left[{1\over 2}(z^2+w^2)\partial_z\partial_w
+{1\over 6}(d_z-d_w)^2 -\right.\\
\left.
-{1\over 12}(z^2-w^2)\left(
\partial_w{z+w\over z-w}\partial_z+
\partial_z{z+w\over z-w}\partial_w\right)
-{1\over 3}(z-w)^2\partial_z\partial_w\right]
\ee

An interesting property of this equation is its locality in
the coordinate space: it does not contain operators like
$1/\partial$. The most singular (when $z\to w$) term reminds
2D fermionic propagator.
Notice that this expression is not translational
invariant. Indeed, the translational invariance is the commutativity
with the operator $s^+=\partial_z+\partial_w$. In the $XXZ$ case,
this operator is deformed (see Appendix), and thus $XXZ$ hamiltonian
is not translational-invariant\footnote{To avoid
confusion, we remind that $H_{XXZ}$ does not
commute with $U_q(sl(2))$. Indeed, it was obtained from the
$U_q(sl(2))$-invariant expression by symmetrization.}

\section*{Acknowledgements.}

I would like to thank A.~Gorsky and S.~Lukyanov for discussions.
I am very indepted to Prof.~A.~LeClair for his kind hospitality in
Cornell University, where the part of this work was done.
The work was partially supported by RFFI Grant No. 97-02-19085.

\section*{Appendix.}

The ${\cal U}_q (\widehat{sl}(2))$-algebra is defined by the generators

\begin{equation}
e_0,\;e_1,\;f_0,\;f_1,\;t_0,\;t_1
\end{equation}

with the commutation relations:

\begin{equation}
\begin{array}{c}
t_ie_jt_i^{-1}=q^{a_{ij}}e_j\\
t_if_jt_i^{-1}=q^{-a_{ij}}f_j\\
{}[e_i,f_j]=\delta_{ij}\frac{t_i-t_i^{-1}}{q-q^{-1}}
\end{array}
\end{equation}

where
\begin{equation}
a_{ij}=\left(\begin{array}{cc} 2&-2\\ -2&2 \end{array}\right)
\end{equation}

and the Serre relations:

\begin{equation}
x_i^3 x_j-[3]x_i^2x_jx_i+[3]x_ix_jx_i^2-x_jx_i^3=0
\end{equation}

where $i\neq j$ and $x=e,\;f$.

The "quantum adjoint action":

\begin{equation}
\mbox{ad}_e(A)=\sum\limits_i e^iAs(e_i)
\end{equation}

 where

\begin{equation}
\sum\limits_i e^i\otimes e_i = \Delta (e)
\end{equation}

The comultiplication in ${\cal U}_q(\widehat{sl}(2))$ is
\footnote{Our definition of $e$, $f$, $t$ differs from the one
in \cite{JM} by $e_i={e^{[JM]}}_i {t^{[JM]}}_i^{-1/2}$,
$f_i={t^{[JM]}}_i^{1/2} {f^{[JM]}}_i$, $t_i=t^{(JM)}_i$} :

\begin{equation}
\begin{array}{c}
\Delta(t_i)=t_i\otimes t_i\\
\Delta(x_i)=x_i\otimes t_i^{-1/2}+t_i^{1/2}\otimes x_i
\end{array}
\end{equation}

The other comultiplication, $\Delta'$, differs from this one
by permuting tensor multipliers: $\Delta'=P\Delta P$.
This is also homomorphism from the universal enveloping to its tensor
square, since this fact obviously does not depend on the order of
tensor multipliers.

The antipode is:

\begin{equation}
\begin{array}{c}
a(t_i)=t_i^{-1}\\
a(e_i)=-q^{-1} e_i\\
a(f_i)=-q f_i
\end{array}
\end{equation}

where again $x_i=e_i,\;f_i$.

We can rewrite the Serre relations as:

\begin{equation}
\mbox{ad}_{x_i}^{1-a_{ij}} (x_j)=0
\end{equation}

and $e_0$ is the vector operator with respect to the
${\cal U}_q(sl(2))$ subalgebra $(e_1,f_1,t_1)$.

We will need the evaluation representation of this algebra, which
is defined as follows. Let $\rho$ be some (finite-dimensional)
representation of ${\cal U}_q (sl(2))$. Then
the evaluation representation $\rho_{\zeta}$ of
${\cal U}_q(\widehat{sl}(2))$ depends on some parameter $\zeta$,
and the space of this representation coincides with the space of
$\rho$ :

\begin{equation}
\begin{array}{c}
\rho_{\zeta}(x_1)=\rho(x_1),\; x=e,f\\
\rho_{\zeta}(e_0)=\zeta\rho(f_1)\\
\rho_{\zeta}(f_0)={1\over \zeta}\rho(e_1)\\
\rho_{\zeta}(t_0)=(\rho_{\zeta}(t_1))^{-1}=(\rho(t_1))^{-1}
\end{array}
\end{equation}

The interesting question is what is the tensor product of two
such evaluation representations. In general it is some
irreducible representation, depending on two parameters.
But for the special values of $\zeta$'s this product turns out
to be reducible.

\underline{Example.}
Consider the product of two spin-$1/2$ representations, with
the parameters $z$ and $w$. Consider the vector
$\downarrow\otimes\downarrow$. Acting on this vector by
$e_1$, we get:

\begin{equation}\label{e1}
\begin{array}{c}
(q^{-{\sigma^3\over 2}}\otimes \sigma^++
\sigma^+\otimes q^{\sigma^3\over 2})
\downarrow\otimes\downarrow=\\=
q^{1/2}\downarrow\otimes\uparrow+q^{-1/2}\uparrow\otimes\downarrow
\end{array}
\end{equation}

And acting on it by $f_0$, we get:

\begin{equation}\label{f0}
\begin{array}{c}
(q^{\sigma^3\over 2}\otimes \sigma^+/z+
\sigma^+/w\otimes q^{-{\sigma^3\over 2}})
\downarrow\otimes\downarrow=\\=
(q^{-1/2}/z)\downarrow\otimes\uparrow+
(q^{1/2}/w)\uparrow\otimes\downarrow
\end{array}
\end{equation}

These two vectors are different at general values of
$z$ and $w$, but when $w=q^2z$  (\ref{f0}) equals
(\ref{e1}) multiplied by ${1\over qz}$.

This means, that at this special value there is a 3-dimensional
subrepresentation in this 4-dimensional tensor product.

In general,

\begin{equation}
\rho_{\zeta}^{1/2}\otimes \rho_{q^2\zeta}^{1/2}\otimes\cdots
\otimes \rho_{q^{2(n-1)}\zeta}^{1/2}\supset \rho_{q^{n-1}\zeta}^{n/2}
\end{equation}


\end{document}